\documentstyle[12pt]{article}

\begin{document}
\baselineskip 0.7cm

\renewcommand{\thefootnote}{\fnsymbol{footnote}}
\setcounter{footnote}{1}

\begin{titlepage}

\begin{flushright}
TU-460
\\
June, 1994
\end{flushright}

\vskip 0.35cm
\begin{center}
{\large \bf
Vector-like Strong Coupling theory with small S and T parameters
}
\vskip 1.2cm
Nobuhiro Maekawa
\footnote
{Fellow of the Japan Society for the Promotion of Science.}
\footnote{
N. Maekawa will be in SLAC in this summer.}
\vskip 0.4cm

{\it Department of Physics, Tohoku University,\\
     Sendai 980, Japan}

\vskip 1.5cm

\abstract{
We propose a mechanism which can reduce the Peskin and Takeuchi's
S, T and U parameters
in dynamical electroweak symmetry breaking models.
It is interesting that not only S but also T parameter can become
small
even if there exists large isospin violation in fermion condensation.
For example, when we take the $SU(2)_L\times U(1)_Y$ breaking mass of
up-type fermion $m_U=1$ TeV and that of down-type $m_D=0$,
we get S$\sim 0.001 N$ and T$\sim 0.05 N$ for the $SU(2)_L\times
U(1)_Y$ invariant masses $M=10$ TeV.
The point is that these parameters are suppressed
by $SU(2)_L\times U(1)_Y$ invariant masses which the vector-like
fermions can have.
}

\end{center}
\end{titlepage}

\newcommand{\tr}{{\rm tr}}
%
%
%
%
Dynamical electroweak symmetry breaking is one of the most attractive
mechanism
which can solve the naturalness problem in the standard model
\cite{TC}.
However recent precision measurements on S and T parameters
give these models rather severe constraints
\cite{STU}.
S parameter gives a severe constraint to the number of new $SU(2)_L$
doublets, which limits the dynamical models like `Walking Technicolor'
\cite{WTC}.
T parameter makes it difficult to break isospin symmetry in dynamical
breaking sector
\cite{chiv}
in spite of the fact that the Nature has very large
isospin violation such as $m_b/m_t<<1$.
Top condensation
\cite{top}
 is attractive as a point that
no new particle exists.
But unfortunately it requires strong fine tuning
\footnote{
This is because the T parameter must be made small.
In order to decrease the composite scale $\Lambda$ very heavy
top quark is needed
\cite{top,bms}.
},
therefore it does not solve the naturalness problem.
If one introduces 4-th generation fermion condensation
\cite{four},
one should introduce degenerate masses of the 4-th up and down-type
quarks,
which is not so natural because the Nature has large isospin violation.

S, T and U parameters are non-decoupling parameters in terms of
$SU(2)_L\times U(1)_Y$ breaking masses $m$. On the other hand,
they must be suppressed by $SU(2)_L\times U(1)_Y$ invariant masses $M$
in the limit $M\rightarrow \infty$ because of decoupling theorem
\cite{AC}.
Therefore even if one introduces new particles which have the
$SU(2)_L\times U(1)_Y$ breaking masses $m$, S, T and U parameters
can be
reduced if they have also $SU(2)_L\times U(1)_Y$ invariant masses
$M>>m$.
By these facts, we can expect that
if the condensation of massive vector-like fields
( i.e. they have $SU(2)_L\times U(1)_Y$ invariant masses $M$ )
\cite{PW,DP,GK,Banks} can
break the standard gauge symmetry,
S, T and U parameters can become small at least in the limit $M>>m$.
In this paper, we would like to discuss dynamical electroweak
symmetry breaking
by vector-like fields' condensation which is induced mainly by
4-fermi interactions.

First, let us discuss the dynamical symmetry breaking by
vector-like fields.
Here we introduce 2 pairs of vector-like fields $Q_{L,R}=
(Q^U,Q^D)_{L,R}=(2,Y)_{L,R}$ and
$U_{R,L}=(1,Y+{1\over 2})_{R,L}$,
in which the numbers are quantum numbers of $SU(2)_L$ and $U(1)_Y$
respectively, and $L$ and $R$ represent the chirality.
Here we adopt the following lagrangian;
\begin{eqnarray}
{\cal L}_4&=&\bar Q (iD^\mu \gamma_\mu -M_Q) Q+\bar U
(iD^\mu \gamma_\mu-M_U) U
             \nonumber \\
   && +({G\over N}(\bar Q {(1+\gamma_5)\over 2} U)(\bar U
{(1-\gamma_5)\over 2} Q)
        +h.c.),
\end{eqnarray}
in which $M_Q$ and $M_U$ are $SU(2)_L\times U(1)_Y$ invariant masses,
$D^\mu$ is a covariant derivative of the standard gauge groups,
and
we neglected every 4-fermi interaction except that in Eq. (1)
for simplicity.
Here we only assume the chiral structure of the 4-fermi interaction,
which we will discuss later.
By using auxiliary field method,
this lagrangian is rewritten like
\begin{eqnarray}
{\cal L}_Y=\bar Q (iD^\mu \gamma_\mu -M_Q) Q+\bar U
(iD^\mu \gamma_\mu-M_U) U
      -{N\over G} \phi^\dagger \phi
      +(\bar Q {(1+\gamma_5)\over 2} U \phi+h.c.).
\end{eqnarray}
We integrate the fermion fields, and get the $1/N$ leading potential
\begin{eqnarray}
V&=&{N\over G}v^2-{N\over 8\pi} I+const,\\
I&=&{\Lambda^4\over 2}[ \ln (1+2\alpha+x_Qx_U)+2\alpha \nonumber \\
 && -(\alpha+\beta)^2\ln {1+\alpha+\beta\over \alpha+\beta}
    -(\alpha-\beta)^2\ln {1+\alpha-\beta\over \alpha-\beta} ] \\
 &\sim& \left\{ \begin{array}{l} {1\over 2}\Lambda^4(1- x_Qx_U)
\ln v^2\quad
                                     (v\rightarrow \infty) \\
                                f(x_Q,x_U) \Lambda^2 v^2+const,
\quad (v\rightarrow 0)
                                 \end{array}\right. \\
x_{(Q,U)}&=&{M_{(Q,U)}^2\over \Lambda^2},\\
\alpha&=&{1\over 2}(x_Q+x_U+v^2/\Lambda^2),\\
\beta&=&\sqrt{\alpha^2-x_Qx_U},\\
f(x,y)&=&{1\over 2(1+x)(1+y)}+{1\over 2}
       -\left\{{x^2\over 2(x-y)}(2\ln \left({1+x\over x}\right)
-{1\over 1+x})
         +(x\leftrightarrow y)\right\}
\end{eqnarray}
with the vacuum expectation value of Higgs field $\langle \phi
\rangle=(v,0)$
and the cut off $\Lambda$.
For any fixed value of $M_U$ and $M_Q$, there exists a critical
coupling
$G_c(x_Q,x_U)=8\pi^2/(\Lambda^2f(x_Q,x_U))$ (see Fig.1).
On the contrary, for any fixed value of $G$ greater than
$8\pi^2/\Lambda^2$,
the critical line exists in $(M_Q, M_U)$ plane, at which $v$ drops
to zero.
This is intuitively understandable.
If the interaction is so strong
that the binding energy becomes larger than sum of the bare masses,
the mass square of this bound state becomes negative and the
symmetric vacuum becomes unstable.

The breaking mass of the fermion $m_U$, the W boson mass $m_W$ and
the Higgs mass $m_H$ can be roughly estimated by
the following relations;
\begin{eqnarray}
m_U&=&v, \\
m_W^2&=&{1\over 2}g_2^2 Z_\phi v^2,\\
m_H^2&=&{1\over 2 Z_\phi}{d^2 V\over dv^2},
\end{eqnarray}
where the renormalization constant $Z_\phi$ and the curvature of
the potential are
\begin{eqnarray}
Z_\phi&=&{N\over 16\pi^2}\left( \ln{1+x\over x}-{1\over 1+x}
          -{1\over 3(1+x)^2} \right), \\
{d^2 V\over dv^2}&=&{Nv^2\over 2 \pi^2}\left( \ln {1+x\over x}
-{1\over 1+x}
           -{1\over 2(1+x)^2}-{1\over 3(1+x)^3}
           \right).
\end{eqnarray}
Here $x=M^2/\Lambda^2$ and we have used the stationary condition
$dV/dv=0$ and the approximation $v^2<<M^2, \Lambda^2$.
{}From the above relations we can easily find that
$\sqrt{N} m_U \sim \sqrt{N} m_H\sim O$(TeV) (see Fig.2).
In order to suppress S and T parameters, we should take
$m_U<<M$, which requires a kind of fine tuning (see Fig.3).
But this is not so strong fine tuning as the top condensation, and
the condensation of the massive vector-like fields is possible under
4-fermi interaction.

Secondly, we would like to estimate the S and T parameters in a
theory with
massive vector-like fields
\cite{vectorST,LS}
in order to see that they are actually suppressed by the $SU(2)_L\times
U(1)_Y$ invariant masses $M$.
For completeness  we introduce one more pair of  vector-like fields
$D_{R,L}=(1,Y-{1\over 2})_{R,L}$.
The mass matrices are taken as
\begin{eqnarray}
M_u=\left( \begin{array}{cc} M_Q & m_U \\ m_U & M_U \end{array}
\right),\
M_d=\left( \begin{array}{cc} M_Q & m_D \\ m_D & M_D \end{array}
\right),\
\end{eqnarray}
in which $M_Q,\ M_U$ and $M_D$ are gauge invariant masses and
$m_U$ and $m_D$ are $SU(2)_L\times U(1)_Y$ breaking masses.
Here we took these mass matrices symmetric for simplicity.
And the fermion mass part of the lagrangian is
\begin{eqnarray}
{\cal L}_M&=&
(\bar Q^U_L, \bar U_L) M_u \left(\begin{array}{c} Q^U_R \\ U_R
\end{array} \right)
+(\bar Q^D_L, \bar D_L) M_d \left(\begin{array}{c} Q^D_R \\ D_R
\end{array} \right)
+h.c. \\
          &=&(\bar U_{L1}, \bar U_{L2})
             \left(\begin{array}{cc} m_{U1} & 0 \\ 0 & m_{U2}
\end{array}\right)
             \left(\begin{array}{c} U_{R1} \\ U_{R2} \end{array}
\right) \\
           && +(\bar D_{L1}, \bar D_{L2})
             \left(\begin{array}{cc} m_{D1} & 0 \\ 0 & m_{D2}
\end{array}\right)
             \left(\begin{array}{c} D_{R1} \\ D_{R2} \end{array}
\right)+h.c.
\end{eqnarray}
in which
\begin{eqnarray}
\left(\begin{array}{c} U_{(L,R)1} \\ U_{(L,R)2} \end{array}\right)=
\left(\begin{array}{cc} c & -s \\ s & c \end{array}\right)
\left(\begin{array}{c} Q^U_{(L,R)} \\ U_{(L,R)} \end{array} \right),\
\left(\begin{array}{c} D_{(L,R)1} \\ D_{(L,R)2} \end{array}\right)=
\left(\begin{array}{cc} \bar c & -\bar s \\ \bar s & \bar c
\end{array}\right)
\left(\begin{array}{c} Q^D_{(L,R)} \\ D_{(L,R)} \end{array} \right).
\end{eqnarray}
The S and T parameters are estimated via the fermion loops
as follows;
\begin{eqnarray}
{\rm S}&=&{N\over 6 \pi}[-4Y(c^2 \ln m_{U1}^2+s^2\ln m_{U2}^2
-\bar c^2 \ln m_{D1}^2
-\bar s^2 \ln m_{D2}^2) \nonumber \\
&&-c^2s^2(6\chi(m_{U1},m_{U2})+{m_{U1}^2+m_{U2}^2\over m_{U1}m_{U2}}
-2) \\
&&-\bar c^2\bar s^2(6\chi(m_{D1},m_{D2})
  +{m_{D1}^2+m_{D2}^2\over m_{D1}m_{D2}}-2)],\nonumber \\
{\rm T}&=&{N\over 8\pi \sin^2 \theta_W m_W^2}[
c^2\bar c^2\theta (m_{U1},m_{D1})+c^2\bar s^2\theta (m_{U1},m_{D2})
\nonumber \\
&&+s^2\bar c^2\theta (m_{U2},m_{D1})+s^2\bar s^2\theta
(m_{U2},m_{D2}) \\
&&-c^2s^2\theta (m_{U1},m_{U2})-\bar c^2\bar s^2\theta
(m_{D1},m_{D2})],\nonumber
\end{eqnarray}
where the functions
\cite{LS}
\begin{eqnarray}
\chi(x,y)&=&{5\over 9} - {4x^2y^2\over 3(x^2-y^2)^2}
-{x^6+y^6-3x^2y^2(x^2+y^2)
          \over 3(x^2-y^2)^3}\ln {x^2\over y^2} \nonumber \\
&&+xy\left[-{1\over 6x^2}-{1\over 6y^2}+{x^2+y^2\over (x^2-y^2)^2}
      -{2x^2y^2\over (x^2-y^2)^3}\ln {x^2\over y^2}\right], \\
\theta(x,y)&=&x^2+y^2-{2x^2y^2\over x^2-y^2}\ln {x^2\over y^2}
              +2xy\left[{x^2+y^2\over x^2-y^2}\ln{x^2\over y^2}
-2\right]
\end{eqnarray}
are non-negative, and zero only if $x=y$.
And $N$ means the number of $SU(2)_L$ doublets.
In the following, we take $m_D=0$, i.e. the isospin is maximally
violated
\footnote{
It is interesting that there exists the region that S parameter
can be negative
when $M\sim m_U$, $m_D=0$ and $Y=-1/2$ (lepton like),
though the mass scale should be rather small ($\sim 100$ GeV)
in order to make T parameter small.
}.
If $M=M_Q=M_U=M_D>>m=m_U$,
then the S and T parameters can be expanded by $m/M$
\begin{eqnarray}
{\rm S}&=&{2N\over 3\pi}(Y+{33\over 60})\left(m\over M\right)^2
    +O\left(\left({m\over M}\right)^4\right), \\
{\rm T}&=&{N m^2\over 40\pi \sin^2\theta_W m_W^2}\left(\left(
m\over M\right)^2
    +O\left(\left({m\over M}\right)^4\right)\right).
\end{eqnarray}
It is sure that the decoupling theorem works.
For example, if we take $m=1$ TeV and $M=10$ TeV,
then S$\sim 0.001 N$ and T$\sim 0.05 N$.
Notice that the parameter T is fairly small
in spite of such a large isospin violation ($m_U=1$ TeV and $m_D=0$).
This may explain the large isospin violation in the Nature.

Finally we would like to discuss what models can realize the above
scenario.
By using the mechanism in this paper, the top condensation is naturally
extended to 4-th family and a anti-family scenario.
This model is interesting because it does not need so strong
fine tuning
and large isospin violation can be realized.
However we would like to discuss another model here.
We introduce
 one  anti-family techni-fermion in addition to
ordinary one family techni-fermion and 2 extended technicolor (ETC)
gauge groups
$SU(N_{TC}+3)_G\times SU(N_{TC})_{AG}$.
Namely, we assign quantum numbers $(N_{TC}+3, 1)$ to the
techni-fermions
and ordinary matters, and $(1, N_{TC})$ to the anti-techni-fermions.
We assume the following breaking pattern;
\begin{eqnarray*}
&& SU(N_{TC}+3)_G\times SU(N_{TC})_{AG}\times G_{SM} \\
&\Lambda_1 &  \nonumber \\
&\rightarrow&SU(N_{TC}+2)_G\times SU(N_{TC})_{AG}\times G_{SM}\\
&\Lambda_2&\nonumber \\
&\rightarrow&SU(N_{TC}+1)_G\times SU(N_{TC})_{AG}\times G_{SM} \\
&\Lambda_3& \nonumber  \\
&\rightarrow&SU(N_{TC})_G\times SU(N_{TC})_{AG}\times G_{SM} \\
&\Lambda & \nonumber  \\
&\rightarrow&SU(N_{TC})_{V}\times G_{SM}\\
&\Lambda_W & \nonumber  \\
&\rightarrow&SU(N_{TC})_{V}\times SU(3)_C\times U(1)_Q,
\end{eqnarray*}
where $G_{SM}=SU(3)_C\times SU(2)_L\times U(1)_Y$ and
$SU(N_{TC})_{V}$ is a vector-like technicolor group.
Under the scale $\Lambda$ vector-like techni-fermions can have
$G_{SM}$ invariant masses.
It is natural to expect that the generation gauge coupling $g_G$ is
larger
than the anti-generation gauge coupling $g_{AG}$ because
$N_{TC}+3>N_{TC}$.
Suppose that $g_G$ is enough strong
at the scale $\Lambda$
to induce the strong chiral 4-fermi interaction like in the
previous discussion.
The 4-fermi interaction induced by the extended technicolor
interactions
\footnote{
In this paper S and T parameters are estimated only via
the fermion loops,
though we should estimate contributions from the bound states.
If 4-fermi interactions are induced by extended technicolor
interactions,
there must exist vector resonances in general.
If the masses of the vector bound states are light,
S and T parameters will be seriously changed.
Our estimation in this paper seems to suggest
that the masses of the vector states
remain heavy, $O(M)$ in spite of the strong 4-fermi interaction.
This may be because in vector channel quadratic divergence does not
exist.
More study is needed at this point.
}
may be below the critical value,
but if the 4-fermi interaction is near critical and
the effect of the strong technicolor interaction $SU(N_{TC})_V$
are also taken into account,
it is not so unnatural to expect that
the condensation becomes possible,
which breaks $SU(2)_L\times U(1)_Y$
against the vacuum alignment by the invariant mass terms
\footnote{
In this model, Witten-Vafa's theorem
\cite{VW} cannot be applied
because chiral structure plays important role to break the electroweak
symmetry.
}.
Except the fact that the techni-fermions are vector-like and have
$SU(2)_L\times U(1)_Y$ invariant masses,
this is effectively so-called gauged Nambu-Jona-Lasinio model
\cite{GNJL},
which can play important roles to solve FCNC problem, light pseudo
Nambu Goldstone problem and mass hierarchy problem.

In conclusion, if vector-like fermions condensate and break the
standard
gauge symmetry, S, T parameters can be small in spite of the large
isospin
violation.
The point is that S, T parameters are suppressed
by $SU(2)_L\times U(1)_Y$ invariant masses which vector-like
fermions can have.
Since the Nature has large isospin violation, we think that this
mechanism is
important to build realistic models.

The author thanks to T. Yanagida for reading this manuscript and
valuable comments.
We would like to thank also D. Suematsu for letting me know
some papers\cite{GK,Banks}.
This work is supported in part by the Japan Society for the
Promotion of
Science.


\newcommand{\journal}[4]{{\sl #1} {\bf #2} {(#3)} {#4}}
\newcommand{\PL}{\sl Phys. Lett.}
\newcommand{\PR}{\sl Phys. Rev.}
\newcommand{\PRL}{\sl Phys. Rev. Lett.}
\newcommand{\NP}{\sl Nucl. Phys.}
\newcommand{\PTP}{\sl Prog. Theor. Phys.}

{\large\bf Figure Caption}



{\bf Fig.1}

\noindent
The inverse of critical coupling $G_c(0,0)/G_c(x,x)=f(x,x)$ with
$x=M^2/\Lambda^2$.

{\bf Fig.2}

\noindent
$SU(2)_Y\times U(1)_Y$ breaking mass of fermion $m_U$ (solid line) and
mass of Higgs $m_H$ (dashed line) with $x=M^2/\Lambda^2$.
In the limit $\Lambda\rightarrow \infty$, i.e. $x\rightarrow 0$,
the ratio $m_H/m_U$ becomes 2, which is so called Nambu-Jona-Lasinio
relation.

{\bf Fig.3}

\noindent
The potential with fine tuning $\hat v=v/\Lambda<<1$.
Here we take $M^2=0.1 \Lambda^2$ and $G=1.636 G_c(0,0)$.

\end{document}